\documentclass[aps, twocolumn, showpacs]{revtex4}
\begin{document}
\title{Double prism evanescent mode experiment: superluminal photons?}
\author { S. C. Tiwari\\
Institute of Natural Philosophy\\
c/o 1 Kusum Kutir Mahamanapuri,Varanasi 221005, India}
\begin{abstract}
A critique on the interpretation of the double prism experiment on the
evanescent microwave modes (arXiv:0708.0681) is presented. It is argued
that these experiments do not give any evidence of superluminal photons.
A physical mechanism based on the inhomogeneous waves of classical electromagnetic
fields is proposed.
\end{abstract}
\maketitle

Recent paper \cite{1} claims that evanescent microwave modes violate special relativity,
and that these modes represent virtual photons of QED. Nimtz and his collaborators have
been doing extensive work on the tunneling experiments since the first report by
Enders and Nimtz in 1992. A critical review can be found in my monograph on superluminal
phenomena \cite{2}. Is there any new evidence for superluminal photons in Nimtz and
Stahlhofen (N-S) recent work \cite{1} ? The aim of the present note is to address this 
question recapitulating relevant points made in \cite{2}. Further it is argued that invoking virtual photons and QED in this context is superfluous. Physics of inhomogeneous plane waves 
and classical electrodynamics are suggested to account for the results of \cite{1}.

In the monograph \cite{2} superluminal electromagnetic (EM) phenomenon is classified
into three catagories: light pulse propagation in material media, special forms of light
beams, and evanescent waves. In Sec.(3.6) of \cite{2} tunneling time and evanescent modes
are discussed in detail. Salient points are given below along with remarks on the
recent N-S work in each case.

[1]-- Sommerfeld drew analogy between evanescent modes and quantum tunneling in \cite{3};
later Hartman's model for quantum tunneling of a particle across a potential barrier and
mathematical similarity between time-independent Schroedinger equation and Helmholtz
equation for the monochromatic EM wave motivated microwave (mw) tunneling experiments. Note
that this formal similarity does not entail physical equivalence: $\psi$ in Schroedinger
equation is the probability amplitude in the Copenhagen interpretation-markedly different
than the Schroedinger's original mechanischer feldskalar, while EM field is a classical
field. Not only this, there are intricate problems in defining Schroedinger-like
description for photon \cite{4}. It is intriguing that Nimtz and Enders in 1993 paper
realize this:' .. in spite of the formal similarities, one has to keep in mind the
striking differences in the interpretation of a probability amplitude of a single
quantum mechanical process and of classical fields', but in the concluding part 
do not adhere to this caution stating,' The electromagnetic-mode experiment is assumed
to correspond to particle tunneling. Obviously Hartman's model describes the tunneling
of both particle and electromagnetic wave packets'. We have argued that this formal
analogy has been overstretched for evanescent waves and that it is misleading.

Remarks on N-S: Since the quantum analogy itself is questionable, introducing QED
and virtual photons is superfluous. Violation of Einstein's relation, Eq.(2) in \cite{1}
for purely imaginary wave number for evanescent modes is non-existent: phenomenological
complex wave number has to be interpreted such that the imaginary part accounts for the
decaying amplitude not the wave number that appears in the Einstein's relation.

[2]-- Sommerfeld in his book mentions that J. C. Bose first observed evanescent modes
for 20 cm mw using two asphalt prisms (now,more than a century ago!). Symmetric double prism experiment of Haibel-Nimtz-Stahlhofen in 2001 was carried out at 9.15 GHz. This is reviewed in \cite{2}.

Remarks on N-S: In this experiment Perspex prisms are used for mw at 3.28 cm same
as that for 2001 experiment, and similar results are obtained. There is no new
result except the inference of zero tunneling time at one meter scale.

[3]-- Nimtz and Haibel in 2000 presented a nice discussion on the conceptual problems
associated with the tunneling experiments. In double prism experiments the tunneling
time is separated into two parts, see Eq.(3.84) in \cite{2}. Time delay due to Goos-Hanchen
shift is interpreted to correspond to the propagating wave with real wave number. The
second part is due to the motion perpendicular to the surface for the imaginary wave
number. Nimtz and Haibel rightly note that the Sommerfeld-Brillouin theory has not been
developed for the evanescent modes. This should mean that the notion of velocity for
these modes a la S-B theory is not clear. What then is the meaning of signal velocity
for evanescent modes? Though recourse is taken to the signal theory in communication
engineering to define the delay times, its relationship with the signal velocity and
tunneling time remains obscure-thus superluminal signals do not follow.

Remarks on N-S: Eq.(1) in \cite{1} defines time delay, and it is assumed to imply that
the real part of the wave vector being zero for the evanescent mode the corresponding 
propagation is also zero. This reasoning is erroneous: evanescent mode is not a 
propagating wave and the imaginary wave number determines the amplitude factor not the
phase. Perhaps a vague similarity could be with the standing wave in which the phase is
constant for all space at each instant of time, and there is no finite propagation velocity.
The claim of superluminal signals is unfounded.

Preceding discussion underlines the importance of the double prism experimental results,
however physical interpretation in terms of superluminal signals is refuted. We argue that
the new twist invoking QED and Feynman diagrams for the evanescent modes would bring
avoidable weirdness to the classical phenomenon of total reflection. Physical significance
attached to the imaginary wave number via Eqs. (1) and (2) in \cite{1} is not correct as
pointed out above. Perhaps it may be of interest to recall that Feinberg in 1967 proposed
Klein-Gordon equation with imaginary mass for tachyon, however Robinett's analysis \cite{5}
raised doubts whether the propagation speed of such a particle was superluminal, see p.114
in \cite{2}. Note that for the imaginary wave number Eq.(2) in \cite{1} setting
$E=mc^2$ leads to imaginary mass for evanescent modes. Such analogies are unphysical.

How do we understand the reported results? An outline of an approach based on 
classical EM theory is presented in this note. Snell's law of refraction shows that for
a light wave propagating from a dense medium to a rare medium the light propagates
along the tangent to the interface boundary of the two media if the angle of incidence
$\theta _i$ is equal to the critical angle. For $\theta _i$ greater than the critical
angle there is total reflection.  Assuming the interface to be the xy-plane in
Cartesian system using simplifying notation the transmitted wave is proportional to
$e^{i(kx-\omega t)}e^{-\gamma z}$. This wave represents a propagating wave along x-direction
in the plane of incidence with exponentially decreasing amplitude factor along the normal
to the plane, i.e. z-direction. Textbooks show that there is no energy flow across
the interface but the boundary conditions on the EM fields require nonvanishing transmitted
fields in the second (rare) medium. This discussion is based on three assumptions \cite{6}:
infinite interface boundary, infinite wave-fronts, and the stationary fields. The effect
of finite sized beams (e.g. Goos-Hanchen shift) on total reflection is reviewed by H. K. V.
Lotsch in the journal Optik (1970), see footnote on page 49 in \cite{6}. The third assumption
is a crucial one for tunneling process. Recall that in an inhomogeneous plane wave the
equi-phase and equi-amplitude planes are in different directions. The monograph by Clemmow
\cite{7} is an important work on this subject; here we follow the short discussion given
on page 83 in \cite{2}.

To motivate the proposed mechanism first we discuss a well known situation: a charge $Q$
is brought into a charge free region. After initial transients an electrostatic Coulomb
field surrounding the charge is established. Now introduce a test charge $q$ anywhere
in the space, it instantly experiences the Coulomb force due to the source charge $Q$.
Similarly suppose a dc voltage is suddenly impressed upon an ideal infinite transmission 
line then the line attains this voltage while the wave-front travels with uniform velocity
$v$ which is determined by the inductance and capacitance per unit length of the line.
There is uniform voltage along the line upto the point $vt$ at any time $t$ after the
application of the voltage. We suggest this analogy for the stationary evanescent fields.
After initial time delay the fields acquire steady state values in the second medium.
What we call tunneling is then the detection of these pre-existing fields which extend to 
infinity though with extremely low amplitudes. The efficiency of the detector and 
the establishment of the steady state are the sole criteria to observe the evanescent modes
instantly. Since the microscopic theory of the interface is not considered here
the mathematical analysis based on classical EM theory can be developed assuming
sources in the Maxwell equations. Let us assume that the total reflection entails
on an average a uniform charge motion along the interface with the velocity less than the
velocity of light. This could be viewed as a current source on the boundary and the
bound fields are calculated as indicated in Eqs. (3.69) to (3.72) in \cite{2}. These
bound fields are evanescent. Thus the tunneling experiment can be explained using
classical theory without invoking superluminal signals.

It may be asked: Do faster-than-light objects exist? Is special relativity sacrosanct?
I believe these questions require thorough discussion on the foundations of space and
time in natural science, see \cite{2} and a recent monograph \cite{8}. Our conclusion
is that both the constancy of the velocity of light and it being a limiting velocity
in nature could at best be considered as convenient conventions, but are devoid of
physical reality.

 The Library facility at Banaras Hindu University is acknowledged.

\end{document}